\documentclass[intlimits,twoside,a4paper]{article}

\usepackage{amsmath,amssymb}
\usepackage{graphicx}

\usepackage[T2A]{fontenc}
\usepackage[cp1251]{inputenc}

\usepackage{epsfig}

\usepackage{color}
\usepackage{wrapfig}

\usepackage{cmpj2}

\issue{2015}{18}{3}{33707}
\doinumber{10.5488/CMP.18.33707}

\title[Renormalized energy of ground and first excited state of Fr\"{o}hlich polaron]%
{Renormalized energy of ground and first excited state of Fr\"{o}hlich
polaron in the range \\ of weak coupling}
\author[M.V.~Tkach \textsl{et al.}]{M.V.~Tkach\thanks{E-mail: ktf@chnu.edu.ua}\,,
Ju.O.~Seti, O.M.~Voitsekhivska, O.Yu.~Pytiuk}
\address{Chernivtsi National University, 2 Kotsyubinsky St., 58012 Chernivtsi, Ukraine}

\date{Received June 10, 2015, in final form July 2, 2015}
\authorcopyright{M.V.~Tkach, Ju.O.~Seti, O.M.~Voitsekhivska, O.Yu.~Pytiuk, 2015}

\begin{document}

\maketitle

\begin{abstract}
Partial summing of infinite range of diagrams for the
two-phonon mass operator of polaron described by Fr\"{o}hlich
Hamiltonian is performed using the Feynman-Pines diagram
technique. Renormalized spectral parameters of ground and
first excited (phonon repeat) polaron state are accurately
calculated for a weak electron-phonon coupling at $T=0$~K.  It is
shown that the stronger electron-phonon interaction shifts the
energy of both states into low-energy region of the spectra. The
ground state stays stationary and the excited one decays at a
bigger coupling constant.
\keywords polaron, phonon, electron-phonon interaction, Green's function, mass operator

\pacs 71.38.-k, 63.20.kd, 63.20.dk, 72.10.Di
\end{abstract}

\section{Introduction}

The concept of polaron, as electron interacting with the
polarization vibrations of the crystal, introduced by Landau in
1933 \cite{Lan33, Lan48} has attracted a permanent attention lately. For a long
time, the theory of polaronic phenomena was developed in the
framework of different physical models using various mathematical
approaches \cite{Bog49, Lee53}. Following the Fr\"{o}hlich's introduction of the Hamiltonian of
electron interacting with nondispersive (optical) phonons of a
dielectric medium via its polarization, in representation of
second quantization \cite{Fro54}, the methods of quantum field theory were
used to solve the polaron problems \cite{Fey55, Whi65, Hae65}. Studying the
renormalized energy of ground state and the effective mass of
structureless Fr\"{o}hlich polaron, three ranges for electron-phonon
coupling were established. The mobility, impedance and optical
conductivity were investigated in detail for these ranges.

Almost fifty years ago J.~Devreese with colleagues \cite{Dev64, Kar69} were
investigating the polaron complexes related to the excited states
of electron-phonon system. In particular, it was established that in the regimes of intermediate and strong coupling, the so-called relaxed excited states (RES) exist \cite{Dev64, Kar69, Dev71,Dev72, Pee83} in the region of energies a little bit bigger than the renormalized energy of the ground polaron state plus the energy of one phonon while the Franck-Condon (FC) excited states are located higher in the energy scale. The treatment of the structure of the excited quasi-stationary states spectra is a complicated problem, constantly attracting attention of theoretical community. The results obtained within different methods were compared in original papers \cite{Mis00, Mis03, Fil06} and in reviews \cite{Mis05, Dev07, Dev09}.

In the process of theoretical investigation of electron-phonon systems, the physical picture was studied in detail and a new mathematical approach \cite{Mis00, Mis03, Fil06} made it possible to avoid the contradictive results previously obtained in the approximations used in earlier papers. The exact diagrammatic quantum Monte Carlo (DMC) method \cite{Mis00} was used in order to solve the problem of  Fr\"{o}hlich polaron RES and FC states and analyze their dependence on the regime of coupling. In the cited papers it was proven that the one-phonon approximation, used in the early papers of J.~Devreese with colleagues, was not capable of correctly describing the optical conductivity in the limit of strong coupling because, in particular, the energy density for RES and FC states was not correctly defined even at a very small coupling constant ($\alpha = 0.05$). Further, in reference \cite{Fil06}, the main results of reference \cite{Mis00} were confirmed by DMC method for a wide range of the coupling constant. These results correlated well with the ones obtained in the memory function formalism (MFF) and strong coupling expansion (SCE) which assumed the FC principle.

We should mention that DMC method \cite{Mis00, Mis03, Fil06, Mis05} made it possible to establish the properties of optical conductivity in the region of excited states of Fr\"{o}hlich polaron and was used to study the high-temperature superconductivity. In particular, in reference \cite{Mis09} it was proven that electron-phonon interaction, together with magnetic sub-system plays a substantial role in the formation of high-temperature superconductivity.

The search for mathematical approaches to the study of the excited states
of Fr\"{o}hlich polarons with intermediate and strong coupling overshadows
the solution of the same problems in the range of weak
electron-phonon coupling for 3D systems. However, with the
appearance of low dimensional nano-systems (quantum dots, quantum
wires and quantum layers, characterized by weak coupling), where
the difference between electron energies resonates with the energy
of confined optical and interface phonons, the attention payed to the
excited polaron states has grown essentially \cite{Fil12, Dev07, Tka03, Pok02, Gla04, Ben97, Tka14, Tka142}. Now
it is necessary to study the renormalized spectrum of excited
states of 3D polaron in the range of weak coupling.

From the papers \cite{Fro54, Fey55, Whi65, Hae65, Tka03} we know that at $\alpha \ll 1$
in one-phonon approximation for the mass operator, the energy of
polaron ground state ($E$) at $T=0$~K shifts into the low-energy
region due to the electron-phonon interaction. In the vicinity of
$E + \Omega$ energies, where $\Omega$ is the polarization phonon
energy, there is observed a wide peak, related to the bound state of polaron with
one phonon. This fact is clear and coordinates with
physical considerations. However, when the coupling constant
increases, the energy of excited state shifts into the opposite side
of the spectrum, contradicting physical consideration because at
$T=0$~K the virtual phonons are not capable of providing their energy to create a
new quasi-particle. So, it is clear that one-phonon approximation
is not valid for an accurate calculation of mass operator (MO).
Thus, the further approximation was to take into account the two
two-phonon diagrams, proportional to $\alpha ^{2}$ in the polaron
Green's function, besides the one-phonon, proportional to
$\alpha$. Herein, the magnitude of renormalized energy of the ground
state became more accurate (the red shift increased) but the peak
of the energy in the region of the bound state, being located in the
left-hand part respectively to that in one-phonon approximation, shifted
into the high-energy region when $\alpha$ increased.

The abovementioned result brings us to the conclusion that the finite number of diagrams
being taken into account in polaron MO
is insufficient to
obtain the correct physical behavior of the excited polaron state when
the coupling constant varies in a wide range. It is evident that one
has to perform a partial summing of infinite number of diagrams
of the respective order.

In this paper we study the renormalized energies of Fr\"{o}hlich
polaron with weak electron-phonon coupling in such approximation
for the MO, which correctly takes into account partially summed
one- and two-phonon diagrams over all orders of the coupling constant.
The result of this approach is that using the Feynman-Pines
diagram technique we obtained a physically correct conclusion:
both the ground and the first excited polaron states for a system
with weak coupling shift into the low-energy region when $\alpha$
increases.

\section{The Fr\"{o}hlich polaron at $T=0$~K}

We consider polaron as an electron interacting with polarization
phonons described by Fr\"{o}hlich Hamiltonian
\begin{equation} \label{EQ:1_}
H=\sum _{\vec{k}}E_{\vec{k}} a_{\vec{k}}^{+} a_{\vec{k}} +\sum
_{\vec{q}}\Omega _{\vec{q}} \left (b_{\vec{q}}^{+} b_{\vec{q}}
+\frac{1}{2} \right )  +\sum _{\vec{k}\, , \vec{q}}\varphi  \,
(\vec{q})\, a_{\vec{k}+\vec{q}}^{+} a_{\vec{k}} {\rm \;
}\left (b_{\vec{q}} +b_{-\vec{q}}^{+} \right ),
\end{equation}
where
\begin{equation} \label{EQ:2_}
E_{\vec{k}} =E+\frac{\hbar ^{2} k^{2} }{2m}\,,\qquad \Omega _{\vec{q}} =\Omega \,
\, \, \,
\end{equation}
are the energies of electron and optical phonons, respectively,
\begin{equation} \label{EQ:3_}
\varphi (q)=\frac{\ri}{q} \sqrt{\frac{2\pi
\hbar \Omega \alpha }{V} \sqrt{\frac{2\Omega }{m} } }
\end{equation}
--- their binding function expressed within the coupling constant
\begin{equation} \label{EQ:4_}
\alpha =\frac{e^{2} }{\hbar } \left(\frac{1}{\varepsilon _{\infty
} } -\frac{1}{\varepsilon _{0} } \right)\sqrt{\frac{m}{2\Omega } }
=\left(\frac{1}{\varepsilon _{\infty } } -\frac{1}{\varepsilon
_{0} } \right)\sqrt{\frac{m}{m_{0}} \frac{\textrm{Ry}}{\Omega } } \,,
\end{equation}
with $m _{0}$ --- the mass of electron in vacuum and $\textrm{Ry} = 13.6$~eV.

It is well known \cite{Abr12} that at $T=0$~K, the renormalized
electron-phonon spectrum is obtained from the poles of Fourier
image of polaron Green's function which, in its turn, through the
Dyson equation
\begin{equation} \label{EQ:5_}
G\, (\vec{k}, \omega )=\left \{ \hbar \omega  -
E_{\vec{k}} -M\, (\vec{k},\omega ) \right \} ^{-1}
\end{equation}
is related with MO expressed in a diagrammatic
form. An analytical calculation of MO in two first orders
over the powers of the coupling constant $\alpha$ is performed.
According to the rules of Feynman-Pines diagram technique, MO of
the first order over $\alpha$ is defined as
\begin{equation} \label{EQ:6_}
\epsfxsize=0.18\textwidth
\raisebox{-20pt}{\epsfbox{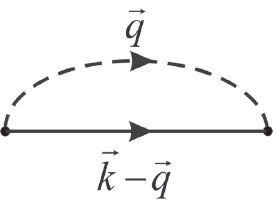}}=M_1(\vec{k},\omega)=\sum_{\vec{q}}\frac{|\varphi(\vec{q})|^2}{\hbar\omega-E_{\vec{k}-\vec{q}}-\Omega+\ri\eta}\,.
\end{equation}

Transiting from summing to integration over $\vec{q}$
and accounting for (\ref{EQ:2_}) and (\ref{EQ:3_}), we obtain
\begin{equation} \label{EQ:7_}
M_{1} (\vec{k},\omega )=\frac{e^{2} \Omega }{4\pi ^{2} }
\left(\frac{1}{\varepsilon _{\infty } } -\frac{1}{\varepsilon _{0}
} \right) \int \frac{\rd^{3} \vec{q}}{q^{2} \left [\hbar \omega
 -E-\frac{\hbar ^{2} }{2m} (\vec{k}-\vec{q})^{2} -
\Omega +\ri\eta \right ]}  \,.
\end{equation}

 Further, it is convenient to introduce the dimensionless MO ${\Large\textsl{m}}= M / \Omega$, with dimensionless energy $\xi$ and
quasi-momentum ($\vec{K}$ and $\vec{Q}$)
\begin{equation} \label{EQ:8_}
\xi =\frac{\hbar \omega -E}{\Omega}\,, \qquad   \vec{K}=\frac{\hbar }{\sqrt{2m\Omega } }\vec{k}\,, \qquad
\vec{Q}=\frac{\hbar }{\sqrt{2m\Omega } } \vec{q}\,.
\end{equation}
In these variables the expression (\ref{EQ:7_}) is rewritten as follows:
\begin{equation} \label{EQ:9_}
{\Large\textsl{m}}_{1} (\vec{K}, \xi )=\frac{\alpha }{2\pi ^{2} } \int_{-\infty }^{\infty } \frac{\rd^{3} \vec{Q}}{Q^{2} \left [\xi
-1-(\vec{K}-\vec{Q})^{2} +\ri\eta \right ]}    \,.
\end{equation}

Integrating in the spherical coordinate system, an exact
analytical expression is obtained
\begin{equation} \label{EQ:10_}
{\Large\textsl{m}}_{1} (K, \xi )=-\frac{\alpha }{K} \, \, \left\{
\begin{array}{ll}
\displaystyle \arctan \left(\frac{K}{\sqrt{1-\xi } }\right ), & \hbox{$\xi \leqslant 1$},  \\[2ex]
\displaystyle \frac{\ri}{2} \ln \left|\frac{\sqrt{\xi -1}
+K}{\sqrt{\xi -1} -K} \right| + \frac{\pi }{2} \theta(K-\sqrt{\xi -1} ), & \hbox{$\xi \geqslant 1$}.
\end{array}\right.
\end{equation}
Herein, at $K=0$
\begin{equation} \label{EQ:11_}
\textrm{Re} {\Large\textsl{m}}_{1} (\xi )=-\frac{\alpha \theta (1-\xi )}{\sqrt{1-\xi } }\,, \qquad \textrm{Im} {\Large\textsl{m}}_{1} (\xi
)=-\frac{\alpha \theta (\xi -1)}{\sqrt{\xi -1} } \,.
\end{equation}

The analysis of MO ${\Large\textsl{m}}_{1}(K,\xi)$  and of the peculiarities of polaron
spectra will be performed further on. Now we should note
that as far as in ${\Large\textsl{m}}_{1}(K,\xi)$ the electron interacts only with
one virtual phonon, it is called a one-phonon MO. Diagram technique
proves that contrary to the two-, three- and $n$-phonon MO,
${\Large\textsl{m}}_{1}(K,\xi)$ is a unitary one, which does not contain an infinite
number of terms. The other terms of complete MO contain an
infinite number of diagrams over all powers of coupling constant.

Now we observe the MO of the second order over the power of
the coupling constant, corresponding to the diagram without an
intersection of phonon lines. According to the rules of diagram
technique, it is written as follows:
\begin{equation} \label{EQ:12_}
M_2^{a}(\vec{k},\omega)=\epsfxsize=0.23\textwidth \raisebox{-20pt}{\epsfbox{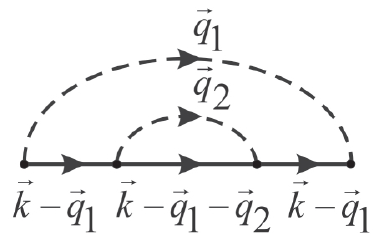}}
=\sum_{\vec{q}_1,\vec{q}_2}\frac{\varphi^2(\vec{q}_1)\varphi^2(\vec{q}_2)}
{\left(\hbar\omega-E_{\vec{k}-\vec{q}_1}-\Omega+\ri\eta\right)^2\left(\hbar\omega-E_{\vec{k}-\vec{q}_1-\vec{q}_2}-2\Omega+\ri\eta\right)}\,.
\end{equation}
Transiting here from summing
to integration, accounting for (\ref{EQ:2_}) and (\ref{EQ:3_}) and using the dimensionless
parameters, we obtain
\begin{equation} \label{EQ:13_}
{\Large\textsl{m}}_{2}^{a} (\vec{K},\xi \, )=\frac{\alpha ^{2} }{4\pi ^{4}
} \int _{-\infty }^{\infty } \frac{\rd^{3} \vec{Q}_{1}
}{Q_{1}^{2} \left [\xi -1-(\vec{K}-\vec{Q}_{1} )^{2} +\ri\eta \right ]^{2} } \int_{-\infty }^{\infty } \frac{\rd^{3}
\vec{Q}_{2} }{Q_{2}^{2} \left [\xi -2-(\vec{K}-\vec{Q}_{1}
-\vec{Q}_{2} )^{2} +\ri\eta \right ]}\,.
\end{equation}

We are going to study the renormalized energies of the bottom of
the ground and excited states of polaron. Thus, here and further
we put $\vec{K}=0$ in order to simplify the analytical
calculations. As a result, the integration in (\ref{EQ:13_}) is performed
exactly and in the region $\xi\leqslant 1$ the expression for
${\Large\textsl{m}}_{2}^{a}(\xi\leqslant 1)$ containing only the real part is obtained in
the following form:
\begin{equation} \label{EQ:14_}
{\Large\textsl{m}}_{2}^{a} (\xi \leqslant 1)=-\frac{\alpha ^{2} }{(1-\xi )^{2} }
\left[\ln \left(1+\sqrt{\frac{1-\xi }{2-\xi } }
\right)-\frac{\sqrt{1-\xi } }{2\left(\sqrt{1-\xi } +\sqrt{2-\xi }\right)} \right].
\end{equation}
Continuing analytically  this expression into the
region $1\leqslant\xi\leqslant2$, one can obtain both the real and the imaginary
parts, while in the region $\xi\geqslant2$, only the real part \cite{Tka03}.

The MO of the second order over the power of the coupling constant,
corresponding to the diagram with the intersection of phonon lines
has the following form:
\begin{eqnarray}
\label{EQ:15_}
M_2^{b}(\vec{k},\omega)&=&\epsfxsize=0.22\textwidth \raisebox{-20pt}{\epsfbox{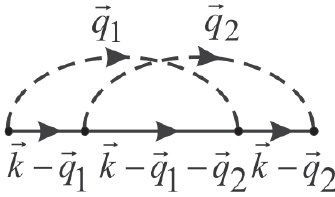}}\nonumber\\
&=&\sum_{\vec{q}_1,\vec{q}_2}\frac{\varphi^2(\vec{q}_1)\varphi^2(\vec{q}_2)}
{\left(\hbar\omega-E_{\vec{k}-\vec{q}_1}-\Omega+\ri\eta\right)^2\left(\hbar\omega-E_{\vec{k}-\vec{q}_1-\vec{q}_2}-2\Omega+\ri\eta\right)
\left(\hbar\omega-E_{\vec{k}-\vec{q}_2}-\Omega+\ri\eta\right)}\,.
\end{eqnarray}
In dimensionless variables it is rewritten as follows:
\begin{eqnarray} \label{EQ:16_}
 {\Large\textsl{m}}_{2}^{b}(\vec{K},\xi )&=&\frac{\alpha ^{2} }{4\pi ^{4} } \int_{-\infty }^{\infty } \frac{\rd^{3} \vec{Q}_{1} }{Q_{1}^{2} \left [\xi -1-(\vec{K}-\vec{Q}_{1} )^{2} +\ri\eta \right ]} \nonumber\\
&&\times \int_{-\infty }^{\infty } \frac{\rd^{3} \vec{Q}_{2} }{Q_{2}^{2} \left [\xi -1-(\vec{K}-\vec{Q}_{2} )^{2} +\ri\eta \right ] \left [\xi -2-(\vec{K}-\vec{Q}_{1} -\vec{Q}_{2} )^{2} +\ri\eta \right ]}  \,.
\end{eqnarray}

At $\vec{K}=0$, integration in (\ref{EQ:16_}) is performed
exactly. As a result, in the region $\xi\leqslant1$ the expression for
${\Large\textsl{m}}_{2}^{b}(\xi\leqslant1)$ containing only the real part is obtained
\begin{equation} \label{EQ:17_}
{\Large\textsl{m}}_{2}^{b} (\xi \leqslant 1)=-\frac{\alpha ^{2} }{(1-\xi )^{2} }
\ln \left[\frac{\sqrt{2-\xi } +\sqrt{1-\xi } }{\xi \sqrt{2-\xi }
+(2-\xi )\sqrt{1-\xi } } \right].
\end{equation}
Continuing analytically this expression into the region
$1\leqslant\xi\leqslant2$, one can obtain both the real and the imaginary parts
while in the region $\xi\geqslant2$, only the real part \cite{Tka03}.

Finally, the functions ${\Large\textsl{m}}_{2}^{a}(\xi\leqslant1)$  and
${\Large\textsl{m}}_{2}^{b}(\xi\leqslant1)$ and their analytical continuations
completely define the MO of the second order over the power of the
coupling constant
\begin{equation} \label{EQ:18_}
\overline{{\Large\textsl{m}}}^{(2)}_{2} (\xi )={\Large\textsl{m}}_{2}^{a}(\xi )+{\Large\textsl{m}}_{2}^{b}(\xi )
\end{equation}
as a complex function of dimensionless energy $\xi$
in the whole range of its variation.

The analytical expressions prove that ${\Large\textsl{m}}_{1}(\xi)$ has a
discontinuity at $\xi=1$ while $\overline{{\Large\textsl{m}}}^{(2)}_{2}(\xi)$~--- at $\xi=1$  and
$\xi=2$.

\begin{wrapfigure}{i}{0.55\textwidth}
\centerline{
\includegraphics[width=0.54\textwidth]{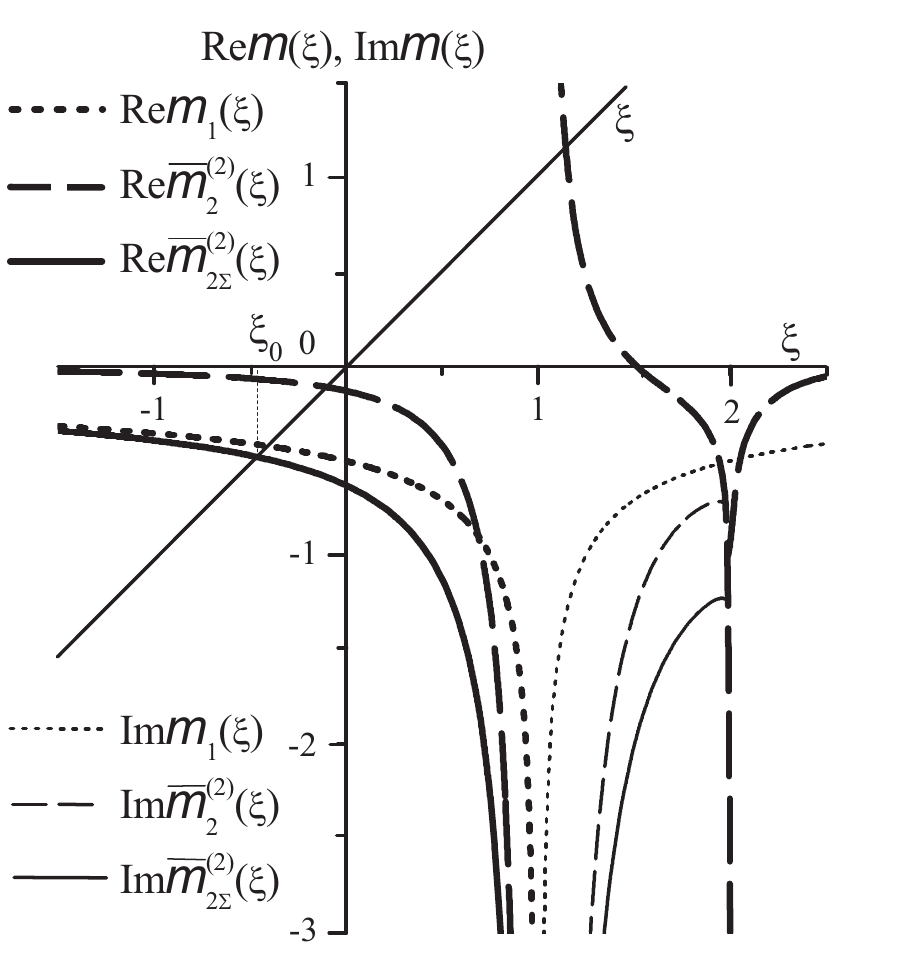}
}
\caption{Dependences of ${\Large\textsl{m}}_1$, $\overline{{\Large\textsl{m}}}_{2}^{(2)}$ and $\overline{{\Large\textsl{m}}}_{2\Sigma}^{2}$ terms on $\xi$ at $\alpha=0.5$.}\label{fig1}
\end{wrapfigure}
In figure~\ref{fig1} typical dependences of MO
$\overline{{\Large\textsl{m}}}_{2\Sigma}^{(2)}(\xi)={\Large\textsl{m}}_{1}(\xi)+\overline{{\Large\textsl{m}}}_{2}^{(2)}(\xi)$
and its terms on $\xi$ calculated at $\alpha=0.5$ are presented.
Here, one can see the properties of MO terms and how they
influence the formation of renormalized energy of the ground
polaron state ($\xi_{0}$).

At $\xi\leqslant1$ the real parts of all terms
[$\textrm{Re} \overline{{\Large\textsl{m}}}_{2\Sigma}^{(2)}(\xi)$] are negative and regularly
decay tending to $-\infty$ at $\xi\rightarrow 1$ from the left. In this region, $\textrm{Im}\overline{{\Large\textsl{m}}}_{2\Sigma}^{(2)}(\xi)=0$, thus, the renormalized energy of polaron ground state is obtained from
the expression $\xi-\overline{{\Large\textsl{m}}}_{2\Sigma}^{(2)}(\xi)=0$. It is
clear from figure~\ref{fig1} that two-phonon
$\overline{{\Large\textsl{m}}}_{2}^{(2)}(\xi)$ being taken into account in addition to the one-phonon ${\Large\textsl{m}}_{1}(\xi)$,
makes $\xi_{0}$ magnitude more precise by increasing its absolute
value.

At $\xi\leqslant 1$, $\textrm{Re} {\Large\textsl{m}}_{1}(\xi)=0$ and $\textrm{Im} {\Large\textsl{m}}_{1}(\xi)<0$, decaying
over the absolute magnitude from $-\infty$ at $\xi=1$ to zero at
$\xi\rightarrow \infty$. In the range $1\leqslant\xi\leqslant2$,
$\textrm{Re}\overline{{\Large\textsl{m}}}_{2}^{(2)}(\xi)$ varies from $\infty$ to $-\infty$
and $\textrm{Im}\overline{{\Large\textsl{m}}}_{2}^{(2)}(\xi)$ varies from $-\infty$ to the
finite negative value. At $2\leqslant\xi$
$\textrm{Im}\overline{{\Large\textsl{m}}}_{2}^{(2)}(\xi)=0$ and $\textrm{Re}\overline{{\Large\textsl{m}}}_{2}^{(2)}(\xi)$
varies from $-\infty$ at $\xi=2$ to zero at $\xi\rightarrow\infty$.

These properties of MO terms determine the features of polaron
spectrum manifested through the dependence of spectral density
$\rho$ on dimensionless energy $\xi$
\begin{equation} \label{EQ:19_}
\rho (\xi )=-2 \textrm{Im} g(\xi )=-\frac{2 \textrm{Im} {\Large\textsl{m}}(\xi )}{\left[\xi -\textrm{Re}{\Large\textsl{m}}(\xi )\right]^{2} +[\textrm{Im} {\Large\textsl{m}}(\xi )]^{2} } \,.
\end{equation}

\begin{figure}[!t]
\centerline{
\includegraphics[width=0.7\textwidth]{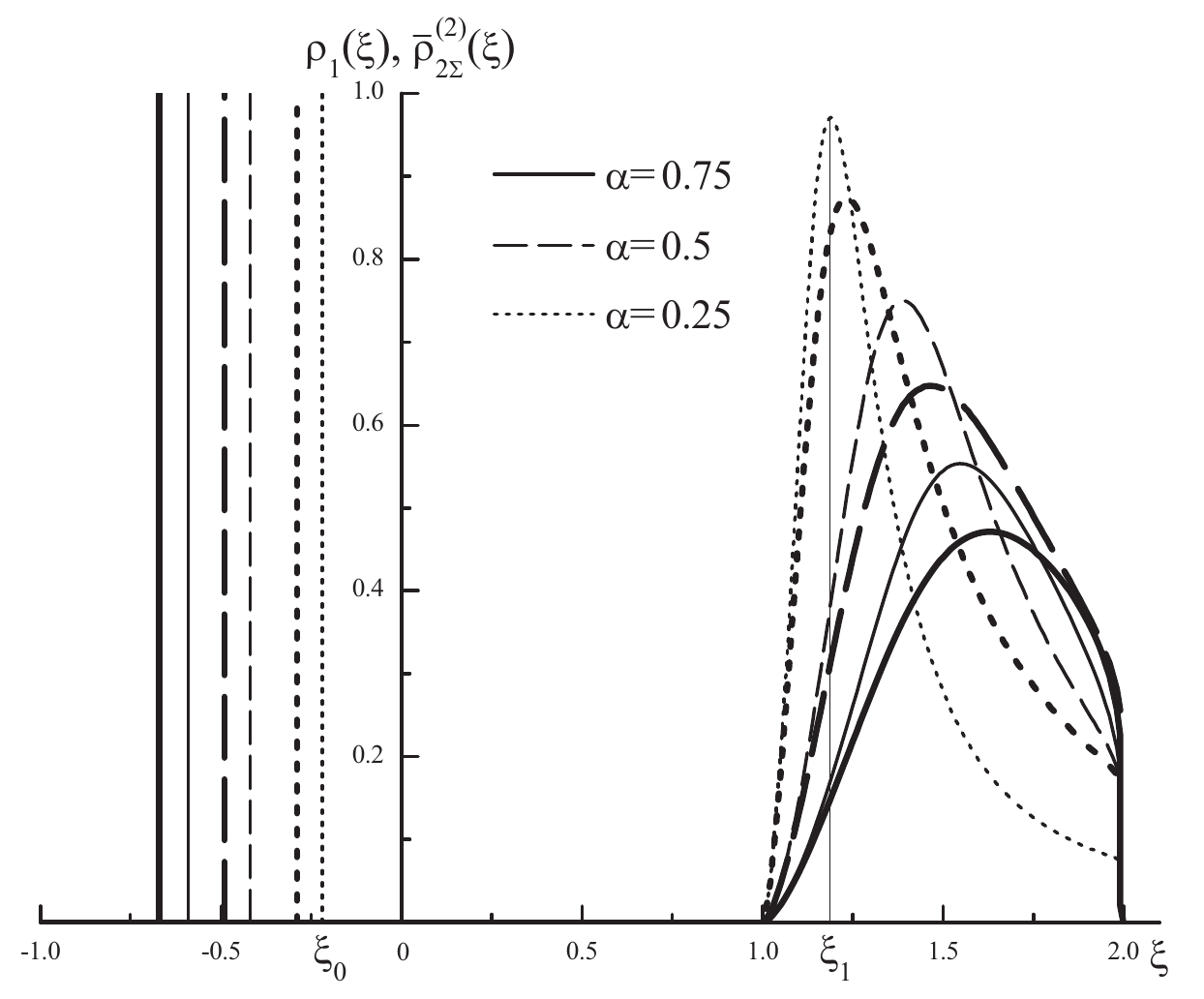}
}
\caption{Dependences of spectral density $\rho$ on energy $\xi$ at different values of the coupling constant $\alpha$ within different approximations for MO. $\rho_{1}$~--- thin curve, $\overline{\rho}_{2\Sigma}^{(2)}$~--- thick curve.}\label{fig2}
\end{figure}

In figure~\ref{fig2}, the function $\rho(\xi)$ is presented at $\alpha=0.25$,
0.5, 0.75 calculated within MO ${\Large\textsl{m}}_{1}(\xi)$ and
$\overline{{\Large\textsl{m}}}_{2\Sigma}^{(2)}(\xi)$. It is clear that independently
of the approximated MO, besides $\delta$-peak corresponding to the
renormalized energy of ground polaron state, the asymmetric peak
of one-phonon repetition with big width ($\gamma$) is observed
because polaron in this state has a small lifetime. At bigger
$\alpha$, the renormalized energy ($\xi_{0}$) shifts into the
negative region while the position of the maximum of one-phonon
repetition ($\xi_{1}$) shifts into the region of higher energies with
the increasing width ($\gamma$) of this peak.

Finally, we should note that one- and two-phonon MO, proportional
to $\alpha$ and $\alpha^{2}$ respectively, cause the
renormalization of polaron ground state energy in such a way that
it shifts into the region of smaller energies when $\alpha$
increases, according to the physical considerations. As for the
excited state, where polaron is bound with one phonon, it is
manifested as asymmetric peak in function $\rho(\xi)$. When
$\alpha$ increases, its maximum shifts into the region of higher
energies, being incorrect from the physical point of view.  In
the next section we show that in order to correctly calculate
the spectral parameters of the excited states, one should
perform a partial summing of infinite ranges of MO diagrams
instead of a finite number thereof.

\section{Ground and first excited polaron states renormalized due to one- and two-phonon processes}

Taking into account the infinite number of diagrams in MO
describing one- and two-phonon processes over all powers of
the coupling constant  but containing no diagram of three- or
more phonons, brings us to the physically correct behavior of the
renormalized energy both of the ground and the first excited state. We refer to such
MO as the two-phonon MO [$M_{2}(\vec{k},\omega)$]. In diagrammatic
representation it is written, according to the rules of
Feynman-Pines diagram technique, as follows:
\begin{equation}
\label{EQ:20_}
\raisebox{120pt}{$M_2(\vec{k},\omega)=\,\,$}\epsfxsize=0.8\textwidth \raisebox{0pt}{\epsfbox{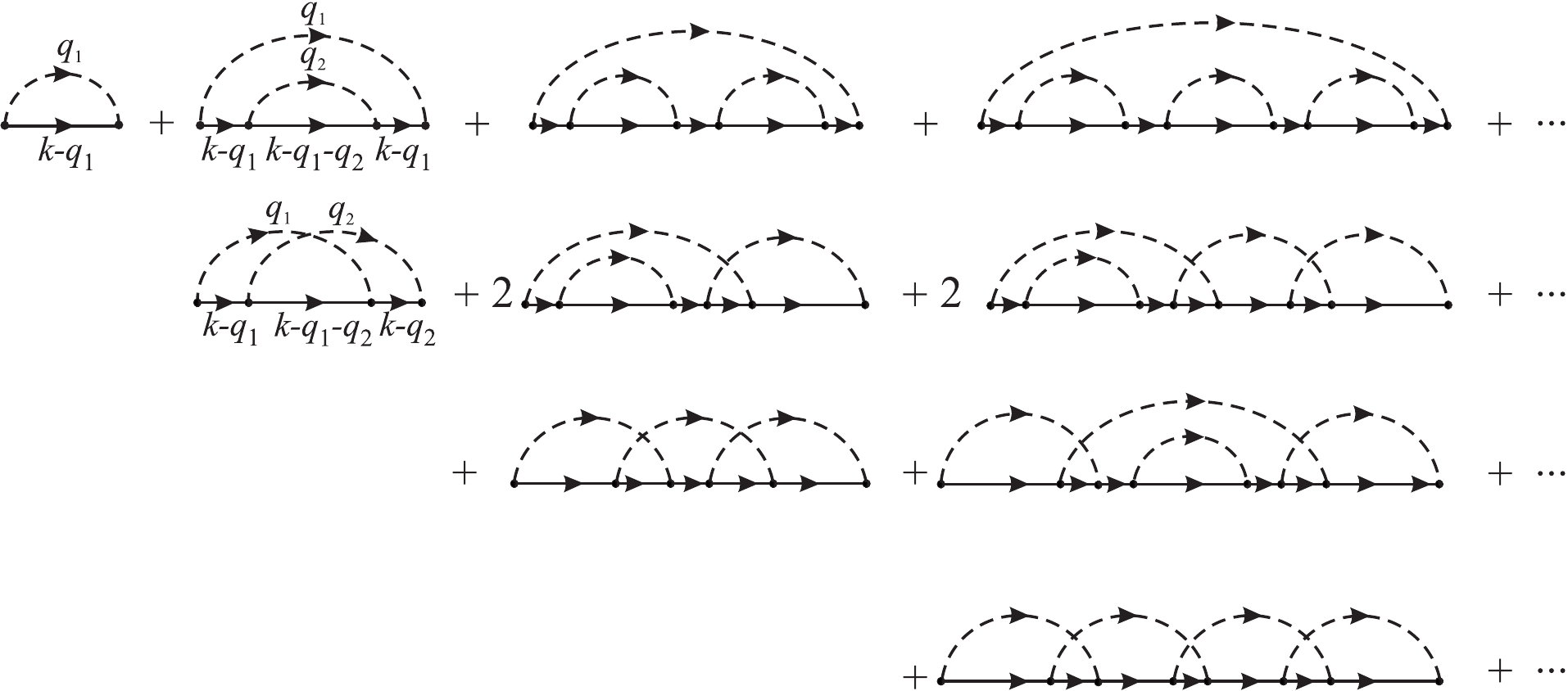}}
\ .
\end{equation}
Performing a complete
partial summing of this range of diagrams, we obtain
\begin{equation} \label{EQ:21_}
 M_{2} (\vec{k},\omega )=\sum _{\vec{q}_{1} }\frac{\left|\varphi (q_{1} )\right|^{2} }{\tilde{\varepsilon }_{\vec{k}-\vec{q}_{1} } } + \sum _{\vec{q}_{1} \vec{q}_{2} }\frac{\left|\varphi (q_{1} )\right|^{2}  \left|\varphi (q_{2} )\right|^{2} }{\tilde{\varepsilon }_{\vec{k}-\vec{q}_{1} } \varepsilon _{\vec{k}-\vec{q}_{1} -\vec{q}_{2} } \tilde{\varepsilon }_{\vec{k}-\vec{q}_{2} } } + \sum _{\vec{q}_{1} \vec{q}_{2} \vec{q}_{3} }\frac{\left|\varphi (q_{1} )\right|^{2}  \left|\varphi (q_{2} )\right|^{2} \left|\varphi (q_{3} )\right|^{2} }{\tilde{\varepsilon }_{\vec{k}-\vec{q}_{1} } \varepsilon _{\vec{k}-\vec{q}_{1} -\vec{q}_{2} } \tilde{\varepsilon }_{\vec{k}-\vec{q}_{2} } \varepsilon _{\vec{k}-\vec{q}_{2} -\vec{q}_{3} } \tilde{\varepsilon }_{\vec{k}-\vec{q}_{3} } } +  ...\, ,
\end{equation}
where
\begin{equation} \label{EQ:22_}
\tilde{\varepsilon }_{\vec{k}-\vec{q}_{1} }=\varepsilon
_{\vec{k}-\vec{q}_{1} } -\sum _{\vec{q}_{2} }\frac{\left|\varphi (q_{2} )\right|^{2} }{\varepsilon
_{\vec{k}-\vec{q}_{1} -\vec{q}_{2} } }=\hbar \omega -E_{\vec{k}-\vec{q}}
-\Omega -\sum _{\vec{q}_{2} }\frac{\left|\varphi (q_{2}
)\right|^{2} }{\hbar \omega -E_{\vec{k}-\vec{q}_{1} -\vec{q}_{2} } -2\Omega }
\end{equation}
is the energetic denominator renormalized due to two-phonon
processes.

Transiting here from summing to integration, integrating over the
angular variables and using the dimensionless parameters (\ref{EQ:8_}), we
obtain an exact analytical expression for a dimensionless
two-phonon MO
\begin{equation} \label{EQ:23_}
{\Large\textsl{m}}_{2} (\vec{K}, \xi )=\sum _{n}{\Large\textsl{m}}_{2}^{(n)} (\vec{K}, \xi) .
\end{equation}

At $K=0$, its terms are as follows:
\begin{align}
\label{EQ:24_}
{\Large\textsl{m}}_{2}^{(1)} (\xi )&=\frac{2\alpha }{\pi } \int_{0}^{\infty}\frac{\rd Q_{1} }{\xi -Q_{1}^{2} -1+\alpha A(\xi ,Q_{1})+\ri \eta } \,, \\
\label{EQ:25_}
{\Large\textsl{m}}_{2}^{(2)} (\xi )&=\frac{\alpha ^{2} }{\pi ^{2} } \int_{0}^{\infty } \int _{0}^{\infty }\frac{\rd Q_{1} }{Q_{1} } \frac{\rd Q_{2} }{Q_{2} } \frac{\displaystyle \ln \left[\frac{2-\xi +(Q_{1}-Q_{2} )^{2} }{2-\xi +(Q_{1} +Q_{2} )^{2} } \right]}{\left [\xi -Q_{1}^{2} -1+\alpha A(\xi, Q_{1} )+\ri\eta \right ] \left [\xi -Q_{2}^{2} -1+\alpha A(\xi, Q_{2} )+\ri\eta \right ]} \,, \\
\label{EQ:26_}
{\Large\textsl{m}}_{2}^{(n)} (\xi )&=4 \left(\frac{\alpha }{2\pi } \right)^{n} \mathop{\int \limits_{0}^{\infty } {}_{\, \ldots}^{(n)} \int \limits_{0}^{\infty } } \frac{\displaystyle \frac{Q_{2}}{Q_{1}} \prod \limits_{s=2}^{n} Q_{s}^{-2} \displaystyle \ln \left[\frac{2-\xi +(Q_{s} -Q_{s-1} )^{2} }{2-\xi +(Q_{s} +Q_{s-1} )^{2} } \right] }{\prod \limits_{s=1}^{n}\left [\xi -Q_{s}^{2} -1+\alpha A(\xi,Q_{s} )+\ri\eta \right ] } \rd Q_{1} \ldots \rd Q_{n}\,,  \qquad (n=2,\, 3,\, \ldots ,\infty ),
\end{align}
where
\begin{equation} \label{EQ:27_}
A(\xi,Q)=\left\{\begin{array}{ll}
\displaystyle \frac{1}{Q} \arctan\left(\frac{Q}{\sqrt{2-\xi } }\right), & \hbox{$\xi \leqslant 2$}, \\[2ex]
\displaystyle \frac{\pi }{Q} \theta(Q-\sqrt{\xi -2})+\frac{\ri}{2Q} \ln \left|\frac{Q+\sqrt{\xi -2}}{Q-\sqrt{\xi -2} } \right|, & \hbox{$\xi \geqslant 2$}.
\end{array}\right.
\end{equation}

The integrals in formulae (\ref{EQ:24_})--(\ref{EQ:26_}) are calculated within
numerical methods. They are typical and contain smooth (without
peculiarities) functions of real variables [numerator in formula
(\ref{EQ:26_})] multiplied by generalized functions $\prod_{s=1}^{n}[
\xi -Q_{s}^{2} -1+\alpha A(\xi, Q_{s} ) + \ri \eta ]^{-1}$.
The presence of the latter causes a different  integration, depending on the range of $\xi$. There are three
specific ranges.

At $\xi\geqslant2$, all factors
$$
\xi -Q_{s}^{2} -1+\frac{\pi \alpha}{Q_s}
\theta \, (Q_{s}-\sqrt{\xi -2} )+\frac{\ri \alpha}{2Q_s} \ln
\left|\frac{Q_{s}+\sqrt{\xi -2}}{Q_{s}-\sqrt{\xi -2}}\right|
$$
are  complex functions, thus, they are all integrals and, hence,
${\Large\textsl{m}}_{2}^{(n)}(\xi\geqslant2)$ is a complex function containing
$\textrm{Re}{\Large\textsl{m}}_{2}^{(n)}(\xi\geqslant2)$ and $\textrm{Im}{\Large\textsl{m}}_{2}^{(n)}(\xi\geqslant2)$ parts.

When $\xi\leqslant 2$, the character of the integrals depends on whether
there exists any solution of the equation
\begin{equation} \label{EQ:28_}
\xi -Q^{2} -1+\frac{\alpha }{Q} \arctan \frac{Q}{\sqrt{2-\xi }}=0
\end{equation}
at fixed $\xi$. At $Q=0$, this equation is rewritten as follows:
\begin{equation} \label{EQ:29_}
\xi +\frac{\alpha }{\sqrt{2-\xi } } =1 ,
\end{equation}
which has an exact real solution
\begin{equation} \label{EQ:30_}
\bar{\xi }=2-\frac{4}{3} \left\{\begin{array}{ll}
\cos ^{2}
\left[\displaystyle \frac{1}{3} \arccos \left(\frac{3\sqrt{3} }{2} \alpha\right)\right], & \hbox{$\displaystyle \alpha \leqslant \frac{2}{3\sqrt{3} }$}, \\[3ex]
\displaystyle \textrm{ch}^{2} \left[\frac{1}{3} \textrm{arcch}\left(\frac{3\sqrt{3} }{2} \alpha \right)\right], & \hbox{$\alpha \geqslant\frac{2}{3\sqrt{3} }$},
\end{array}\right.
\end{equation}
producing two regions for $\xi \leqslant 2$: $\xi\leqslant\overline{\xi}$ and $\overline{\xi}\leqslant\xi\leqslant2$.

At $\xi\leqslant\overline{\xi}$, the equation (\ref{EQ:28_}) does not have any
solution, thus the functions in respective integrals for
${\Large\textsl{m}}_{2}^{(n)}(\xi)$ have no poles and, hence, the MO
${\Large\textsl{m}}_{2}^{(n)}(\xi)$  are real functions.

In the region $\overline{\xi}\leqslant\xi\leqslant2$, the equation (\ref{EQ:28_}) has
the solution in the point $Q_0$ at fixed $\xi$. Thus, the integrals in
${\Large\textsl{m}}_{2}^{(n)}(\xi)$ contain this specific point and are calculated
using  Dirac identity
\begin{equation} \label{EQ:31_}
\int _{0}^{\infty }\frac{\Phi(Q)}{f(Q)+\ri \eta } \rd Q = \mathcal{P} \int _{0}^{\infty}\frac{\Phi(Q)}{f(Q)}\rd Q
-\ri\frac{\pi }{|f\,'(Q)|_{Q=Q_{0} } } \int _{0}^{\infty }\Phi(Q)\delta(Q-Q_{0})\rd Q,
\end{equation}
where $\Phi(Q)$ is the regular function and $f(Q=Q_{0})=0$.

\begin{figure}[!b]
\centerline{
\includegraphics[width=0.55\textwidth]{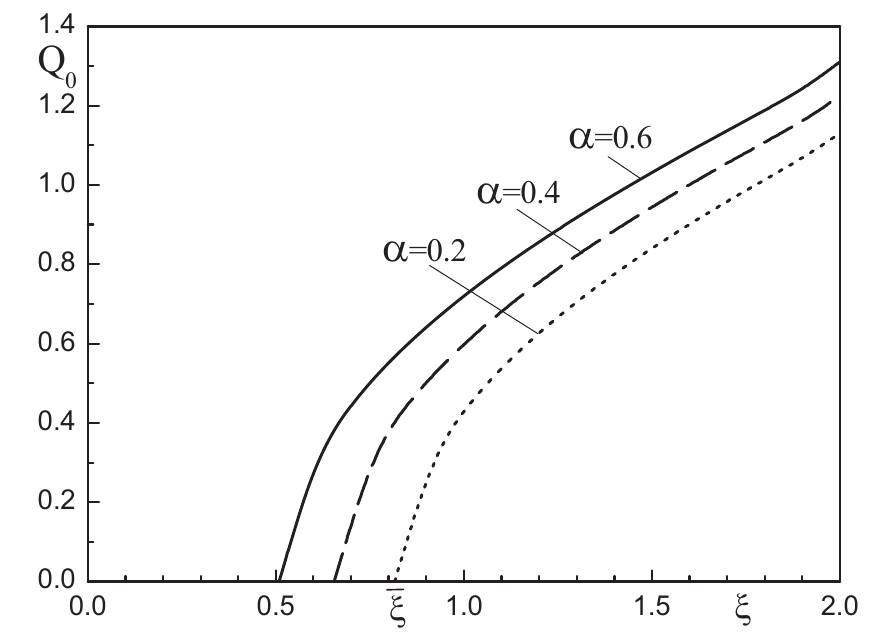}
}
\caption{$Q_{0}$ as function of $\xi$ at $\alpha=0.2$, 0.4, 0.6.}
\label{fig3}
\end{figure}

At $\xi=2$, the equation (\ref{EQ:28_}) has an exact real solution
\begin{equation} \label{EQ:32_}
Q_{0} (\xi =2,\alpha )=\frac{2}{\sqrt{3} }
\left\{\begin{array}{ll}
\displaystyle \cos \left[\frac{1}{3} \arccos
\left(\frac{3\sqrt{3} \pi \alpha }{4} \right)\right], & \hbox{$\displaystyle\alpha \leqslant \frac{4}{3\sqrt{3} \pi}$}, \\[3ex]
\displaystyle \textrm{ch} \left[\frac{1}{3} \textrm{arcch}\left(\frac{3\sqrt{3} \pi \alpha }{4} \right)\right], & \hbox{$\displaystyle\alpha \geqslant
\frac{4}{3\sqrt{3} \pi}$},
\end{array}\right.
\end{equation}
thus, the function $Q_{0}(\xi,\alpha)$ smoothly increases from 0 to
$Q_{0}(2,\alpha)$ in the range $\overline{\xi}\leqslant\xi\leqslant2$, figure~\ref{fig3}.

The presented analysis proves that ${\Large\textsl{m}}_{2}^{(1)}(\xi)$ is given by
an analytical expression
\begin{equation} \label{EQ:33_}
{\Large\textsl{m}}_{2}^{(1)}(\xi )=\frac{2 \alpha }{\pi }
\left\{
\begin{array}{ll}
\displaystyle \phantom{\mathcal{P}}\int\limits_{0}^{\infty }\frac{\rd Q_{1}}{\displaystyle \xi -Q_{1}^{2} -1+\frac{\alpha }{Q_{1} } \arctan \frac{Q_{1} }{\sqrt{2-\xi } } }\,,
& \hbox{$\xi \leqslant \bar{\xi }$},  \\[5ex]
\displaystyle \mathcal{P}\int\limits_{0}^{\infty }\frac{\rd Q_{1}}{\displaystyle \xi-Q_{1}^{2} -1+\frac{\alpha }{Q_{1} } \arctan \frac{Q_{1} }{\sqrt{2-\xi } } }
-\frac{\ri \pi Q_{0} }{\displaystyle \left|3 Q_{0} +1-\xi -\frac{\alpha \sqrt{2-\xi } }{2-\xi +Q_{0}^{2} } \right|}\,,
& \hbox{$\bar{\xi }\leqslant \xi \leqslant 2$},  \\[8ex]
\displaystyle \phantom{\mathcal{P}}\int\limits_{0}^{\infty }\frac{\left[\displaystyle \xi -Q_{1}^{2} -1+\frac{\pi \alpha }{Q_{1} } \theta(Q_{1} -\sqrt{\xi -2})-\frac{\ri \alpha}{2Q_{1} } \ln \left|\frac{Q_{1} +\sqrt{\xi-2} }{Q_{1} -\sqrt{\xi -2} }
\right|\right]\rd Q_{1} }{\displaystyle \left[\xi - Q_{1}^{2}-1+\frac{\pi \alpha }{Q_{1} } \theta(Q_{1} -\sqrt{\xi -2})\right]^{2}
+\frac{\alpha^{2}}{4Q_{1}^{2} } \left[\ln \left|\frac{Q_{1} +\sqrt{\xi -2} }{Q_{1} -\sqrt{\xi -2} } \right|\right]^{2} }\,,
& \hbox{$2\leqslant \xi$}.
\end{array}\right.
\end{equation}


All integrals in ${\Large\textsl{m}}_{2}^{(n)}(\xi)$ are calculated in a similar way.

\begin{wrapfigure}{i}{0.55\textwidth}
\centerline{
\includegraphics[width=0.54\textwidth]{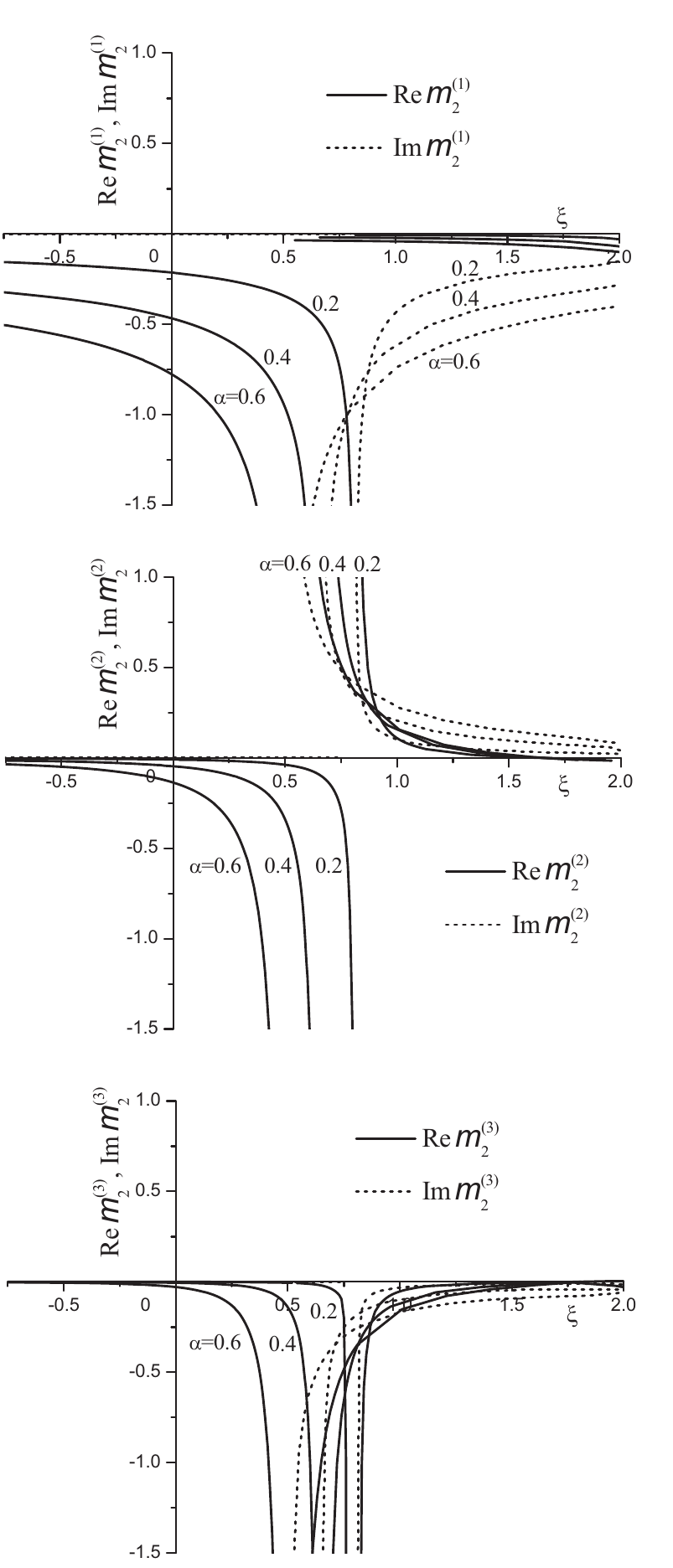} 
}
\caption{MO ${\Large\textsl{m}}_{2}^{(n)}$ as a function of $\xi$ at $\alpha=0.2$, 0.4, 0.6.}
\label{fig4}
\end{wrapfigure}
The first three terms of MO ${\Large\textsl{m}}_{2}^{(n=1,2,3)}(\xi)$ at
$\alpha=0.2$, 0.4, 0.6 are shown in figure~\ref{fig4}. It is clear that at
$\xi\leqslant\overline{\xi}$, $\textrm{Im}{\Large\textsl{m}}_{2}^{(n)}(\xi)=0$ and
$\textrm{Re}{\Large\textsl{m}}_{2}^{(n)}(\xi)<0$ independently of the coupling constant $\alpha$ for all
$n$, herein \linebreak $\lim_{\xi \rightarrow \overline{\xi}} {\Large\textsl{m}}_{2}^{(n)}(\xi) \rightarrow - \infty$. The absolute magnitude $|\textrm{Re}{\Large\textsl{m}}_{2}^{(n)}(\xi)|$ increases
for a bigger coupling constant $\alpha$. In the range
$\overline{\xi}\leqslant\xi\leqslant2$, the real and imaginary parts of
${\Large\textsl{m}}_{2}^{(1)}(\xi)$ and ${\Large\textsl{m}}_{2}^{(3)}(\xi)$ are negative while those
of ${\Large\textsl{m}}_{2}^{(2)}(\xi)$ and ${\Large\textsl{m}}_{2}^{(4)}(\xi)$ are positive.

\begin{figure}
\centerline{
\includegraphics[width=0.84\textwidth]{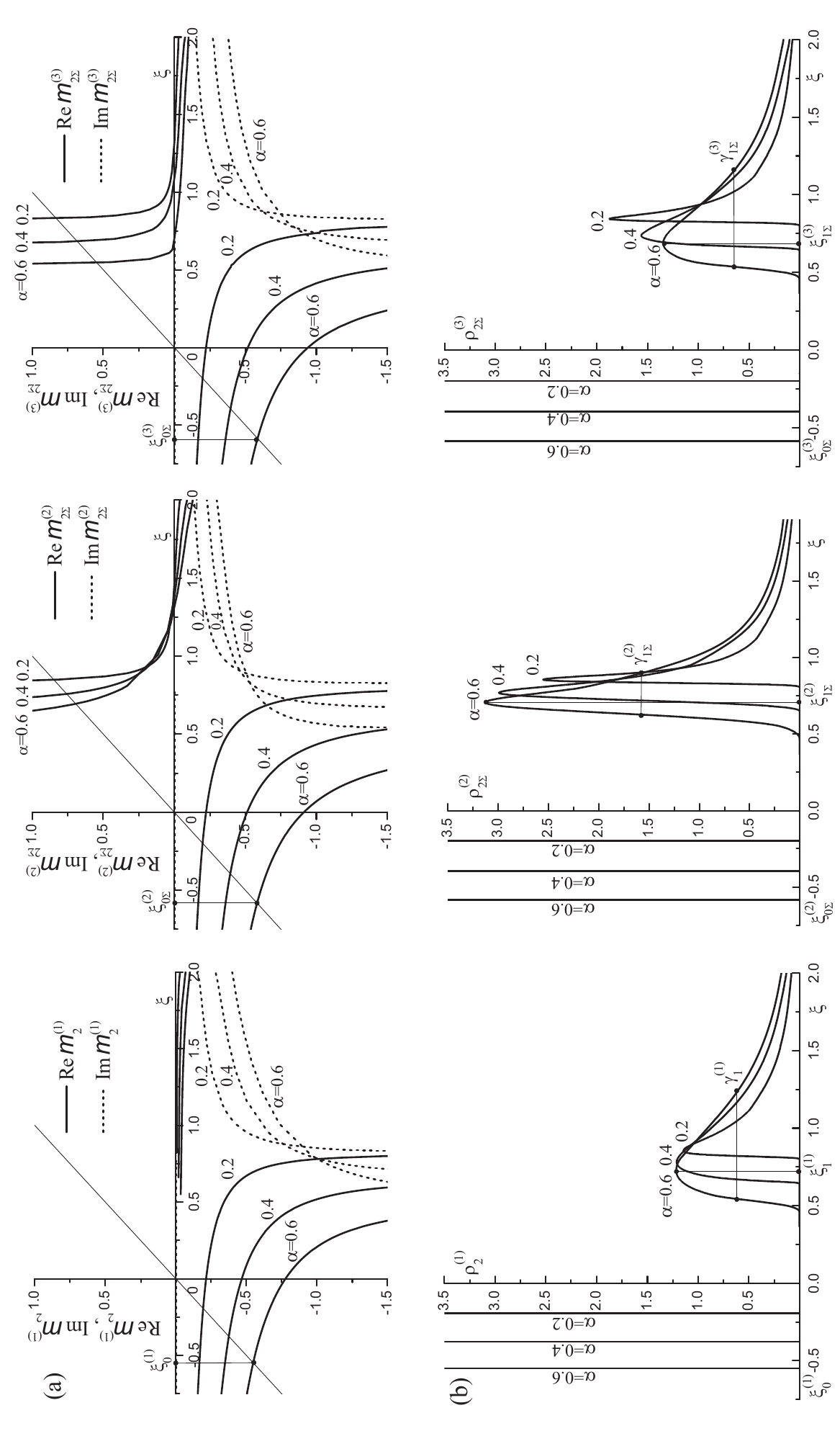}
}
\caption{MO ${\Large\textsl{m}}_{2\Sigma}^{(n)}$ (a) and energy density $\rho_{2\Sigma}^{(n)}$ (b) as functions of $\xi$ at $n=1$, 2, 3 and $\alpha=0.2$, 0.4, 0.6.}
\label{fig5}
\end{figure}

In figure~\ref{fig5}, the functions of real and imaginary parts of MO
${\Large\textsl{m}}_{2\Sigma}^{(n=1,2,3)}(\xi)$ (a) and densities of energies
$\rho_{2\Sigma}^{(n=1,2,3)}(\xi)$ (b) are presented at $\xi\leqslant2$.
The region $\xi>2$ is not observed because here for correct
results one has to take into account the MO renormalized due to the
three-phonon processes. Figure~\ref{fig5}~(a) proves that when
$\xi\leqslant\overline{\xi}$, $\textrm{Im}{\Large\textsl{m}}_{2\Sigma}^{(n)}(\xi\leqslant\overline{\xi})=0$ and, thus, there is no decay
of the renormalized ground state. The real part of MO
$\textrm{Re}{\Large\textsl{m}}_{2\Sigma}^{(n)}(\xi\leqslant\overline{\xi})$ is negative and its
absolute value increases at bigger $\alpha$. Both features bring
us to the fact that the energy density has a $\delta$-like shape
$\rho_{2\Sigma}^{(n)}(\xi)=\delta(\xi-\xi_{0\Sigma}^{(n)})$ with the peak at
the renormalized ground state energy.

In the range $\overline{\xi}\leqslant\xi\leqslant2$, the energy density
$\rho_{2\Sigma}^{(n)}(\xi)$ in all approximations has the shape of
asymmetric quasi-Lorentz peak, arising due to the first excited
state of polaron, interpreted as a bound state of electron with
one phonon. The maximum of $\rho_{2\Sigma}^{(n)}$ in the energy scale
fixes the energy of this bound state and the width
($\gamma_{2\Sigma}^{(n)}$) of the peak at the half of its height
defines the decay of this quasi-stationary state or its lifetime ($\tau_{2\Sigma}^{(n)}=\hslash/\gamma_{2\Sigma}^{(n)}$).

We should note that in spectroscopy, according to the physical
characteristics of renormalized ground and the first excited states,
the respective peaks of the function $\rho_{2}(\xi)$ are referred to as
phononless and one-phonon repetitions. The function $\rho_{2\Sigma}^{(n)}(\xi)$ and its spectral
parameters ($\xi_{0\Sigma}^{(n)}$, $\xi_{1\Sigma}^{(n)}$, $\gamma_{1\Sigma}^{(n)}$) are
shown in figure~\ref{fig5}~(b) for different $\alpha$ and MO approximation
(${\Large\textsl{m}}_{2\Sigma}^{(n)}$).

According to the physical considerations and the behavior of optical conductivity at small $\alpha$ revealed in paper \cite{Mis00}, the energies of both states shift into the low-energy region when the coupling constant increases. Herein, the ground state is a stationary
one (not decaying) and the decay of the excited quasi-stationary state increases.

The spectral parameters also depend on the approximation of MO
${\Large\textsl{m}}_{2\Sigma}^{(n)}$. Figure~\ref{fig5}~(b) and table~\ref{tbl1} prove that
the energies of the ground and the excited states
($\xi_{0\Sigma}^{(n)}$, $\xi_{1\Sigma}^{(n)}$) for bigger $n$ are uniformly
defined more precisely, while the decay
$\gamma_{2\Sigma}^{(n)}$ is defined varying between minimal values
at odd $n$ and maximal values at even ones. From the table~\ref{tbl1} it is clear
that at small $\alpha$, the difference between the energies of the
ground and excited states correlates with the magnitude of the phonon
energy. At a bigger $\alpha$ it increases, which is probably not a physical
property but the result of insufficiency of two-phonon approximation
in MO used for the calculation of renormalized energies.

\begin{table}[!t]
\caption{Dependences of the spectral parameters in the first three orders on the magnitude of $\alpha$.} \label{tbl1} 
\begin{center}
\renewcommand{\arraystretch}{0}
\begin{tabular}{|c||c|c|c|}
\hline {$\alpha$}&
{0.2}& {0.4}& {0.6} \strut\\
\hline
\hline
 {$\xi_{0}^{(1)}$}&  {--0.194}&  {--0.376}&  {--0.548} \strut\\
\hline
 {$\xi_{1}^{(1)}$}&  {0.856}&  {0.780}&  {0.728} \strut\\
\hline
 {$\xi_{1}^{(1)}-\xi_{0}^{(1)}$}& {1.050}&   {1.154}&  {1.276}\strut\\
\hline
 {$\gamma_{1}^{(1)}$}&   {0.247}&   {0.494}&   {0.694} \strut\\
\hline
\hline
 {$\xi_{0\Sigma}^{(2)}$}& {--0.199}&  {--0.395}&  {--0.583} \strut\\
\hline
 {$\xi_{1\Sigma}^{(2)}$}&  {0.854}&  {0.767}&  {0.705}  \strut\\
\hline
 {$\xi_{1\Sigma}^{(2)}-\xi_{0\Sigma}^{(2)}$}&  {1.053}&  {1.162}&  {1.288} \strut\\
\hline
 {$\gamma_{1\Sigma}^{(2)}$}&  {0.077}&  {0.172}&  {0.273} \strut\\
\hline
\hline
 {$\xi_{0\Sigma}^{(3)}$}&  {--0.20}&  {--0.396}&  {--0.585} \strut\\
\hline
 {$\xi_{1\Sigma}^{(3)}$}&  {0.84}&   {0.74}&  {0.69} \strut\\
\hline
 {$\xi_{1\Sigma}^{(3)}-\xi_{0\Sigma}^{(3)}$}&  {1.04}&  {1.136}&  {1.275}  \strut\\
\hline
 {$\gamma_{1\Sigma}^{(3)}$}&  {0.118}&  {0.374}&  {0.594} \strut\\
\hline
\end{tabular}
\renewcommand{\arraystretch}{1}
\end{center}
\end{table}

\section{Conclusions}

Using the Feynman-Pines diagram technique, the exact analytical calculation of MO for the Fourier image of polaron Green’s function is performed in the first and the second order over the electron-phonon coupling constant. It is shown that though such an approximation makes the renormalized energy of polaron ground state more precise, but even at a weak coupling ($\alpha<1$) the energy and decay of the first excited state is evaluated very roughly. Firstly, the peak of the energy density of the first phonon repetition in the energy scale is located higher than the energy $E+\Omega$, however, according to the physical considerations it should be lower because it is produced by the interaction between electron and virtual phonon ($T=0$~K). Secondly, the increasing coupling constant causes the shift of the peak of one-phonon repetition  into the high-energy region, which is not correct either.

A partial summing of all the infinite range of MO diagrams that do
not contain three- and more phonon energies is
performed. The two-phonon MO and the energy density are calculated and
their properties are analyzed for the renormalized ground and
first excited polaron states. It is shown, for the first time, that
according to the physical considerations, the stronger
electron-phonon interaction, i.e., an increasing coupling constant,
shifts the energy of the ground and first excited polaron states into
the low-energy region. Herein, the ground state stays stationary and the decay of the excited state increases. At small
$\alpha$, the difference between the energies of the ground and the
excited polaron states correlates with the magnitude of the phonon
energy.

The developed approach of partial summing of diagrams containing
three-, four- and $n$-phonon processes in polaron MO principally
makes it possible to obtain more accurate renormalized energies and decays
both of the ground state and multi-phonon repetitions. However, an
increasing number of repetitions and the exactness of their spectral
parameters brings us to complicated analytical and numeric
calculations.


\vspace{-5mm}

\ukrainianpart

\title{Перенормована енергія основного і першого збудженого стану полярона Феліха зі слабким зв’язком}
\author{М.В. Ткач, Ю.О. Сеті, О.М. Войцехівська, О.Ю. Питюк}
\address{Чернівецький національний університет ім.~Ю.~Федьковича, \\ вул. Коцюбинського, 2, 58012 Чернівці, Україна}

\makeukrtitle

\begin{abstract}
\tolerance=3000%
Методом діаграмної техніки Фейнмана-Пайнса виконано парціальне підсумовування безмежного ря\-ду діаграм двофононного масового оператора полярона, що описується гамільтоніаном Фрeліха. Перенормовані спектральні параметри основного та першого збудженого поляронного стану (фононного повторення) коректно розраховані для електрон-фононної системи зі слабким зв'язком при $T=0$~К. Показано, що сильніша електрон-фононна взаємодія зміщує енергії обох станів у низькоенергетичну область спектра. Основний стан залишається стаціонарним, а загасання збудженого~--- зростає при збільшенні константи зв'язку.
\keywords полярон, фонон, електрон-фононна взаємодія, функція Гріна, масовий оператор
\end{abstract}

\end{document}